\documentclass[12pt]{iopart}

\usepackage[dvips]{graphicx}
\usepackage{dcolumn}
\usepackage{bm}
\usepackage{subfigure}
\usepackage{hyperref}
\usepackage{array}

\begin{document}

\title[]{Hierarchical structure of stock price fluctuations in financial markets}

\author{Ya-Chun Gao$^{1,3}$, Shi-Min Cai$^{2}$, Bing-Hong Wang$^{3}$}

\address{$^{1}$National Synchrotron Radiation Laboratory,
University of Science and Technology of China, Hefei Anhui, 230026, PR China}
\address{$^{2}$ Web Sciences Center, University of Electronic
Science and Technology of China, Chengdu Sichuan, 611731, PR China}
\address{$^{3}$Department of Modern Physics, University of Science and Technology of China, Hefei
Anhui, 230026, PR China}

\eads{\mailto{kuangben@mail.ustc.edu.cn} (Y.-C. Gao), \mailto{shimin.cai81@gmail.com} (S.-M.Cai), \mailto{bhwang@ustc.edu.cn} (B.-H. Wang)}

\begin{abstract}
The financial market and turbulence have been broadly compared
on account of the same quantitative methods and several
common stylized facts they shared. In this paper, the She-Leveque (SL) hierarchy, proposed to explain the anomalous scaling exponents deviated from Kolmogorov monofractal scaling of
the velocity fluctuation in fluid turbulence, is applied
to study and quantify the hierarchical structure of
stock price fluctuations in financial markets.
We therefore observed certain interesting results: (i)
The hierarchical structure related to multifractal
scaling generally presents in all the stock price fluctuations we investigated.
(ii) The quantitatively statistical parameters that describes
SL hierarchy are different between developed financial markets and
emerging ones, distinctively. (iii) For the high-frequency
stock price fluctuation, the hierarchical structure varies with different time period. All these results provide a novelty
analogy in turbulence and financial market dynamics and a insight to
deeply understand the multifractality in financial markets.
\end{abstract}

\maketitle

\section{Introducation\label{1}}

The financial market has been deemed as a complex system due to the large number of interacting individuals. However, according to the association between the interactions and stock movements, the study of time series of stock prices provides an entry to explore the intrinsic interacting mechanisms of stock markets for experts in scientific community. Bacherlier firstly assumed
the evolution of stock prices as Brownian motion, the
simplest stochastic processes \cite{Bacherlier}, while advanced investigation by Mandelbrot pointed out the returns of cotton prices did not follow the Gaussian distribution, but rather L\'{e}vy stable distribution \cite{Mandelbrot}.
Furthermore, a fatter tail of the return distribution than L\'{e}vy stable distribution was observed by Mantegna and Stanley when analyzing the S$\&$P 500 index
constructed from a series of stock prices. Since then, the \emph{stylized facts}, namely the statistical features of financial time series in global stock markets are
investigated by both economists and physicists\cite{Mantegna1999,Bouchaud,Sornette,Challet}. Several stylized facts are revealed comprehensively and empirically, such as the fat tail in the return distribution and long-range volatility correlation
\cite{Liu,Gopikrishnan,Wang,Zhou2002,Cai2006a,Cai2006b}, and simple agent-based models are successfully raised to reproduce and explain these statistical features \cite{Lux,Challet2000,Tesfatsion,Wang2005,Zhou2007}.

As the statistical techniques to quantitatively measure the time series of stock prices have been commonly adopted in turbulence for a long while, the analogies in the statistical properties between turbulence and stock market dynamics have been surveyed since 1996 by Ghashghaie et al., who presented the similarities in probability densities of foreign exchange rate changes and the velocity difference of turbulence
flow depending on the time delay and spatial separation respectively, as well as intermittency and cascade dynamics \cite{Ghashghaie}.

Further study by Mantegna and Stanley revealed that there was intermittency in both processes, yet the shapes of probability density functions are different, when  they systematically compared the statistical features of the S$\&$P 500 index with velocity of turbulent air \cite{Mantegna1996,Mantegna1997}.
Inspired by these parallel analysis, a multifractal process is available to model the time series of stock price on account of the exhibition of intermittency and cascade dynamics\cite{Muzy2000}, as the multifractal structure has been
extensively found through the structure function \cite{Schmitt2000,Sun2001,Wei2005,Jiang2008}.

Although there are plenty of works in describing the multifractal structure in financial markets, few of them have been engaged in exploiting the intrinsic understanding of this phenomenon.
While in turbulent fluid flows, the analogous anomalous scaling exponents are explained by the SL hierarchy model, which is advanced by She and Leveque after Kolmogorov and his collaborators improving the linear scaling law \cite{She1994,She1995}. In this work, we formulate the SL hierarchy from turbulence to investigate the time series of stock prices from developed to emerging financial markets, and find the hierarchical structure associating with the multifractal scaling generally rooting in them.
The hierarchy provides extended interpretation of the fractal properties in time series of stock prices and simplifies the description with fewer parameters of the multifractal measurement, from which statistical parameters of various values are apparently distinguished from developed to emerging financial markets. Besides, a relevance between the hierarchical structure and different period of financial market is observed in high-frequency time series of stock prices.

In the following section, we first introduce the financial time series to be analyzed,
and then the theory of SL hierarchy. In section 3, we will give plenty of experimental
results on the hierarchical structure of stock price fluctuation in financial market.
Finally, the conclusion is given in the last section.

\section{Materials and Methods \label{2}}

\subsection{Data sets}
The composition of experimental data includes two parts.
One is the daily close prices of 7 stock indices
selected from American, European, and Chinese financial markets \cite{Yahoo}.
Specifically, the stock indices in American financial markets
consist of Dow Jones Industrial Average (DJIA),
Standard $\&$ Poor 500 (S$\&$P500), and Nasdaq composite (Nasdaq),
all of which range from 4st January, 1986 to 6st September, 2011, including the same $6475$ data points. From European financial markets we selected DAX and FTSE 100 indices, the corresponding durations of which start from 11st November, 1990 (DAX) and 6st January, 1986 (FTSE) and both end at 6st September, 2011, leading to the data lengths of $5256$ (DAX) and $6486$ (FTSE), respectively. The remaining two daily indices are Shanghai Composite index (SCI) and Hang Seng index (HSI) of Hong Kong from Chinese financial markets. The duration of SCI
with $2989$ daily stock prices is from 4st January, 2000 to 6st September, 2011, while
HSI with size of $6133$ ranges from 31st December, 1986 to 6st September, 2011.
Besides the daily-frequency time series, we also investigated the minute-to-minute prices of HSI in order to investigate the effect of time changes on the hierarchical structure. The high-frequency time series of $16,6634$ HSI stock prices step over 3 years, from 3rd January, 1994 to 31st December,1996.

\subsection{Method}
For a given time series of stock prices $s(t)$,
the \emph{return} is defined as $r(t,\tau)=s(t+\tau)-s(t)$,
where $\tau$ is time scale. In a range of $\tau$,
the SL hierarchy suggests there is a relationship between moments of different orders,
\begin{equation}
\label{eq1}
[\frac{X_{p+2}(\tau)}{X_{p+1}(\tau)}]=A_{p}[\frac{X_{p+1}(\tau)}{X_{p}(\tau)}]^{\beta}[X^{\infty}(\tau)]^{1-\beta},
\end{equation}
where $X_p(\tau)=\langle |r(t,\tau)|^p \rangle$ is the $p$th order moment of volatility $|r(t,\tau)|$
denoted by $p$th order structure function and
$\langle \cdot \rangle$ indicates the statistical average.
$X^{\infty}(\tau)=\lim_{p \rightarrow \infty} X_{p+1}(\tau)/ X_{p}(\tau)$ suggests that
$X^{\infty}(\tau)$ is dominated by large volatility and therefore describes
the largest amplitude fluctuations in the time series of stock prices.
The parameter $\beta$ characterizing the SL hierarchy is in a interval $(0,1)$
and $A_p$ is proportional coefficient as a function of $p$.

Recalling the scaling law $X_{p}(\tau)\sim \tau^{\xi(p)}$ derived
from the structure function \cite{Muzy2000}, equation (\ref{eq1})
implies the scaling model \cite{She1994,Ching2004}
\begin{equation}
\label{eq2}
\xi(p)=h_0p+C(1-\beta^p),
\end{equation}
where $h_0$ and $C$ are two other parameters of
the SL hierarchical model. According to Eqs. (\ref{eq1}) and (\ref{eq2}),
$X^{\infty}(\tau) \sim \tau^{h_0}$ is obtained. Also, the nonlinear dependence of $\xi(p)$ on $p$ indicates a multifractal scaling, hence the deviation from the
monofractal scaling is mainly characterized by $\beta$ and $C$.
In particular, $\beta$ measures the degree of multifractality, and the monofractal structure of the stock price fluctuation will be achieved
when $\beta \rightarrow 1$. The parameter $C$ initially represents
the codimension of the set of the largest amplitude fluctuation of the
flow for the multifractal description of fluid turbulence \cite{She1994}.
Ching \emph{et al} introduced it for the first to study the heat rate variability
and advanced a comprehension that a larger (smaller) occurring probability of large
amplitude fluctuations is in correspondence with a smaller (larger) value of $C$ \cite{Ching2004}, which is helpful to interpret the stock price fluctuations.

Based on the above theoretical analysis, the procedure to check
whether SL hierarchical structure exists in the stock price variation
is displayed as follows \cite{Ching2002}:

(1) It is assumed that there is a scaling property with
power-law relationship between normalized structure functions
\begin{equation}
\label{eq3}
\frac{X_p(\tau)}{[X_n(\tau)]^{p/n}} \sim \{\frac{X_q(\tau)}{[X_n(\tau)]^{q/n}}\}^{\rho_n(p,q)},
\end{equation}
which is known as the generalized extended self-similarity in fluid turbulence \cite{Benzi1995,Benzi1996}.

(2) In the case of SL hierarchy, the exponent $\rho_n(p,q)$ in Eq. (\ref{eq3}) is connected only with $\beta$ of scaling model,
\begin{equation}
\label{eq4}
\rho_n(p,q)=\frac{n(1-\beta^p)-p(1-\beta^{n})}{n(1-\beta^q)-q(1-\beta^{n})}.
\end{equation}

(3) Fixing the values of $n$ and $q$, and gradually increasing $p$ with step $\delta p=0.2$,
 equation (\ref{eq4}) is transformed into a new fomulation,
\begin{equation}
\label{eq5}
\Delta \rho_n (p+\delta p,q)=\beta^{\delta p}\Delta \rho_n (p,q)-\frac{\delta p(1-\beta^n)(1-\beta^{\delta p})}{n(1-\beta^q)-q(1-\beta^{n})},
\end{equation}
where $\Delta \rho_n (p,q)=\rho_n (p+\delta p,q)-\rho_n (p,q)$. By
plotting $\Delta \rho_n (p+\delta p,q)$ vs $\Delta \rho_n (p,q)$, we can check
the relation of Eq. (\ref{eq5}) and the validity of the SL hierarchy.

\section{Empirical results}\label{3}
\begin{figure}
\center
\includegraphics[width=9cm]{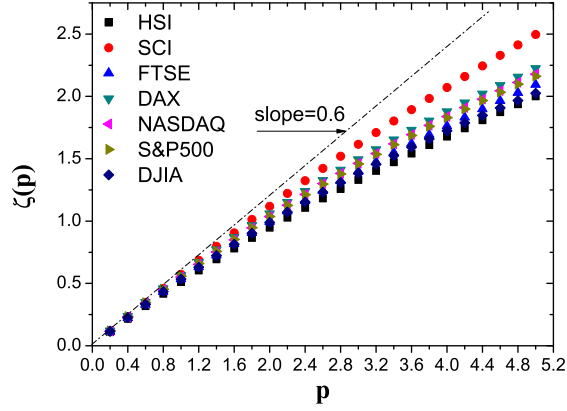}%
\caption{\label{Fig:1} (Color online) The dependence $\xi(p)$ on $p$. The nonlinearly increase of
$\xi(p)$ in respect to $p$ shows the existence of multifractal scaling
and diverse degree of multifractlity for all stock price fluctuations.
The dot-dash line with slope being 0.6 is as a reference of the curves.}
\end{figure}

Before the procedure that verifies the
validity of the SL hierarchy in stock price fluctuations,
we first study the scaling properties of $X_p(\tau) \sim \tau^{\xi(p)}$ with a series of p
and depict the dependence of $\xi(p)$ on $p$. In Fig. \ref{Fig:1},
it is observed that for all stock price fluctuations $\xi(p)$ increase in respect to
$p$ with varying degrees of nonlinearity, especially the curve of SCI is
distinctively different from others. This result suggests
that the multifractal scaling efficiently exists,
yet the degrees of multifractlity is diverse on the basis of the scaling model of Eq (\ref{eq2}).

\begin{figure}
\center
\includegraphics[width=9cm]{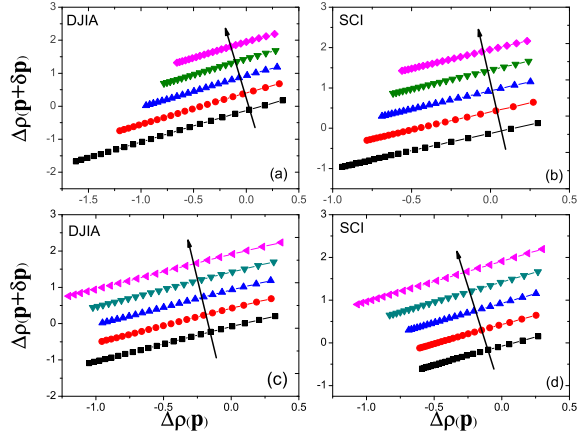}%
\caption{\label{Fig:2}(Color online) The scatter plots of $\Delta \rho_n (p,q)$ vs $\Delta \rho_n (p,q)$ for
stock price fluctuations of DJIA and SCI. We set $\delta p=0.2$ for all plots. Other parameters for
(a) and (b) are $q=1$ and $n=1.6, 1.8, 2, 2.2, 2.4$, while for
(c) and (d) they are $n=2$ and $q=0.6, 0.8, 1, 1.2, 1.4$. These parameters
increase in the direction of the arrow. Note that the slopes of
$\Delta \rho_n (p,q)$ vs $\Delta \rho_n (p,q)$ are evidently insensitive to
the choice of $n$ and $q$.}
\end{figure}

As the theoretical analysis of scaling property is affirmed by the nonlinear relationships between $\xi(p)$ and $p$ presented above, we urge ourselves to further
perform the procedure of Eqs. (\ref{eq3})-(\ref{eq5}) to check the validity
of SL hierarchical structures in the 7 stock price fluctuations. It is achieved through drawing the scatter plots of $\Delta \rho_n (p+\delta p,q)$ vs $\Delta \rho_n (p,q)$, where a series of values of $\Delta \rho_n(p,q)$ can be obtained after exponents of $\rho_n(p,q)$ are estimated from Eq. (\ref{eq3}) with various $p$.
 Fig. \ref{Fig:2} only presents the typical scatter plots of DJIA and SCI with $\delta p=0.2$
because the other 5 scatter plots perform similar features.
The slopes observed from the these scatter plots are in correspondence to the theoretical
analysis of Eq. (\ref{eq5}), which demonstrates that
the SL hierarchical structure indeed roots in stock price fluctuations.
In addition, the scaling property is repeatedly checked by employing a variety of $n$ and $q$, as is shown in Fig. \ref{Fig:2}, frow which we testify that the slopes
of $\Delta \rho_n (p+\delta p,q)$ vs $\Delta \rho_n (p,q)$
are insensitive to the parameters $n$ and $q$. Besides, it should
be noticed that the curves are shifted by tuning the offsets with 0.5.

\begin{figure}
\center
{\includegraphics[width=9cm]{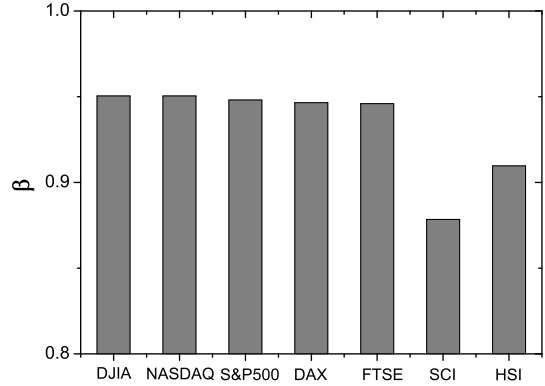}}
\caption{\label{Fig:3} Estimated values of $\beta$ of stock price fluctuations in 7 financial markets.
One can find that they are obviously discriminable in respect to
the development and geographical region of financial markets.}
\end{figure}

Based on the scatter plots of $\Delta \rho_n (p+\delta p,q)$ vs $\Delta \rho_n (p,q)$,
$\beta$ can be estimated from the slopes, in the light of Eq. (\ref{eq5}). In
Fig. \ref{Fig:3}, the values of $\beta$ for all 7 share price volatilities are presented, demonstrating the SL hierarchical structure is compatible with the multifractal scaling. It's quite remarkable from the figure that $\beta$ is discriminable in respect to the development and geographical
region of financial markets, which is reflected in the following details: The developed American and European financial markets
oppose larger values of $\beta$ showing similar SL hierarchical structures of stock price fluctuations
and much more tendency to monofractal scaling. On the contrary, a lower value
of $\beta$ for the emerging Chinese financial market suggests a more regularity of stock price fluctuations that may arising from a market economy that is defective. And then, for Hong Kong financial market, the value of $\beta$ falls in the middle because the geographical region of Hong Kong cause it commonly affected by the Chinese domestic
economy and global economic trend.

To well understand the SL hierarchy, She and Waymire (SW) introduced in
Ref. \cite{She1995} a multiplicative random cascade
that consists of two dynamic components corresponding to scaling modeling
of Eq. (\ref{eq2}). It is illustrated that the first term $h_0 p$ associates with the basic component that generates the singular dynamics cross a continuous scales, while
the second term $C(1-\beta^p)$ correlates with another component that
modulates the singular structure through the multiplication of $\beta$
in discrete steps. Utilizing the relationship
between $\tau$ and $X^{\infty}$ declared in former analysis, it can be inferred that how the terms in Eq. (\ref{eq2})
really affect the scaling model \cite{Ching2004,Ching2002}.

\begin{figure}
\center
\includegraphics[width=9cm]{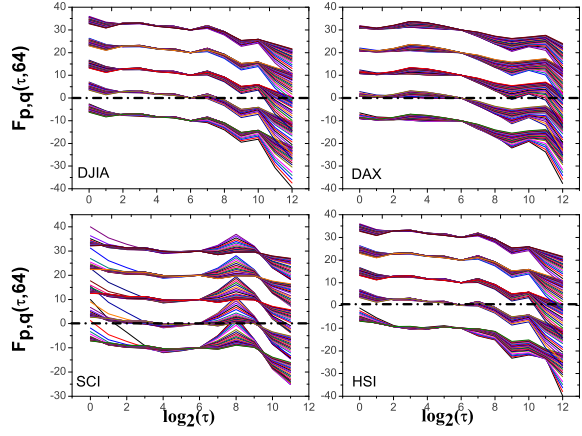}%
\caption{\label{Fig:4} (Color online)
$F_{p,q}(\tau,64)$ vs $\log_2(\tau)$ for various values of $p$ (superimposed) at
a certain choices of $q$. These curves from top to bottom correspond to
$q=1.2, 1.6, 2, 2.4, 2.8$ and are shifted by tuning the offset being 10. Both
of them suggest that the $\tau$-independent $X^{\infty} (\tau)$ exists in a range of $\tau$.}
\end{figure}

Since $X^{\infty}$ cannot be directly calculated from finite $p$, an
alterative method is that we rewrite Eq. (\ref{eq1}) as
\begin{equation}
\label{eq6}
X_{p} \sim [X^{\infty}(\tau)]^{p} \{ \frac{X_{q} (\tau)}{X^{\infty}(\tau)^q}\} ^{\Gamma(p,q)},
\end{equation}
where $\Gamma(p,q)=(1-\beta^p)/(1-\beta^q)$ \cite{Ching2004,Ching2002}.
With the distinct value of $\tau$ and $\tau_0$, Equation (\ref{eq6})
can give rise to a new formulation after some algebra, which is
defined as
\begin{equation}
\label{eq7}
F_{p,q}(\tau,\tau_0)=\log_2[\frac{X^{\infty(\tau)}}{X^{\infty(\tau_0)}}]=\frac{\log_2[X_p(\tau)/X_p(\tau_0)]-\Gamma(p,q)[X_p(\tau)/X_p(\tau_0)]}{p-q \Gamma(p,q)}.
\end{equation}
The existence of the invariant $F_{p,q}(\tau,\tau_0)$ in a range of $\tau$ and $\tau_0$
illustrates that $X^{\infty} (\tau)$ is $\tau$-independent. Herein we set $\tau_0=64$, and plot $F_{p,q}(\tau,\tau_0)$ of DJIA, DAX, SCI and HSI changing with $\tau$ at a linear-log scale respectively, at a set of $p$ with certain choices of $q$, as shown in Fig. \ref{Fig:4}. In this figure, different clusters of
$\log_2(\tau)$-dependent $F_{p,q}(\tau,\tau_0)$ on the basis of different $q$ are shifted by gradually tuning offset of 10,
and the dash-dot line gives the basic line. It is worth noting that $F_{p,q}(\tau,\tau_0)$ is independent of $p$ and $q$, and what is more important, $F_{p,q}(\tau,\tau_0)$ is approximately consistent with zero in a range of $\tau$
(i. e., $\tau$-independent $X^{\infty} (\tau)$ in a range of $\tau$), therefore, the value $h_0 \sim 0$ in Eq. (\ref{eq2})
is statistically ascertained and consequently $\xi(p) \sim C(1-\beta^p)$. Similar results
are also reached for other financial markets.

\begin{figure}
\center
\includegraphics[width=9cm]{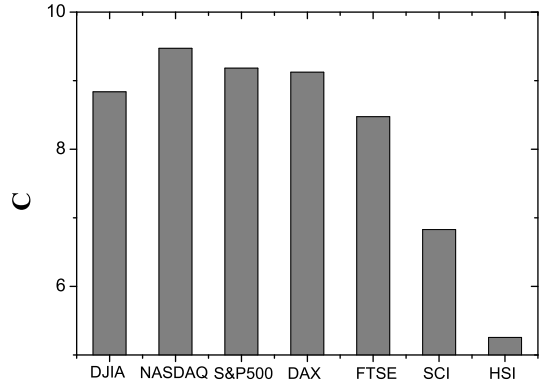}
\caption{\label{Fig:5}Estimated values of $C$ of stock price fluctuations in 7 financial markets.
One can find that these emerging financial markets corresponds to
much lower $C$ comparing with the developed financial markets. This
means that the large stock price fluctuations much easily occur in
emerging financial markets.}
\end{figure}

The aforementioned statement reaches a critical relationship, the parameter $C$ of scaling model can be approximate to
the statistically average of $\xi(p) / (1-\beta^p)$ under the condition of various $p$.
Fig. \ref{Fig:5} shows diverse values of $C$, which are much lower for SCI and HSI
than those in the developed stock markets. In particular,
HSI owns the lowest $C$, indicating a largest occurring probability of large stock price
fluctuation, which is agreed with the empirical observations. As an active banking center, the stock market of Hong Kong is easily impacted by both the geographic region economy and global economic trend, giving rise to a major occurrence of large stock price fluctuations.

So far, the results show the existence of SL hierarchical structure generally in the daily share prices, and for the sake of a deeper insight into the financial market, we turn to explore the high frequency stock price fluctuations. The time series of minute-to-minute HSI stock prices from 1994 to 1996 are analyzed following the same procedure to calculate the parameters
of scaling model, and also find the presence of SL hierarchical property. Furthermore, in Fig. \ref{eq6}, the estimated values of $\beta$ are depicted for one year and total duration, which changes for different year, implying
the SL hierarchical structure varies for different financial period. For example, in the year of 1996 that mostly approaches to Asian financial crisis, the degree of multifractality with the smallest $\beta$ is the most
irregular. On the other hand, we cannot efficiently estimate the parameter $C$ because the evidence of
$\tau$-independent $X^{\infty} (\tau)$ is absent.

\begin{figure}
\center \label{Fig:6}
\includegraphics[width=9cm]{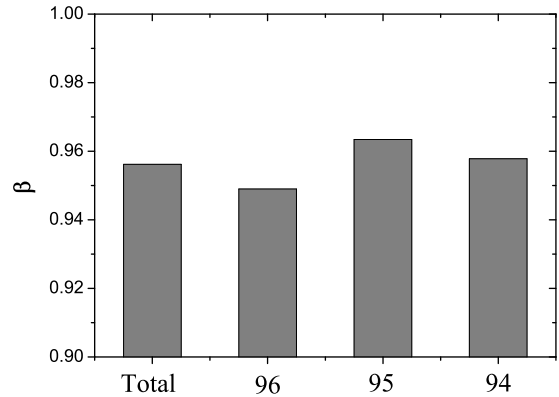}
\caption{Estimated values of $\beta$ of stock price fluctuations for high frequency
HSI from a certain year and total duration. One can find the
the degree of multifractality of stock price fluctuation corresponding to the smallest $\beta$ in the
year 1996 that mostly approaches to Asian financial crisis is the most
irregular.}
\end{figure}

\section{Conclusion}
In conclusion, this work contributed to a new study on stock price fluctuations of 7 developed or emerging financial markets in the first place to adopt the quantitative measurement of SL hierarchy, which was originally projected to characterize the derivation from Kolmogorov monofractal scaling of the velocity fluctuations in fluid turbulence.
According to the three calculated parameters $h_0$, $\beta$ and $C$ of the scaling model, we verified that the SL hierarchical structure which is in connection with multifractal scaling
generally exists in both daily and minute-frequency stock price
fluctuations, considering $0 < \beta < 1$.
However, the degree of multifractality differs between financial markets with different developmental levels and geographical regions for the diverse values
of $\beta$ of their daily stock price volatilities. For instance,
the developed American and European stock markets
corresponding to larger $\beta$ possess a lower degree of multifractality than those emerging financial markets. Moreover, the occurrence of large stock price fluctuations is also smaller for these developed stock markets, taking the higher value of the parameter $C$ into account.
We made further efforts to find that the SL structure alters as the time period of stock price fluctuations changes, by analyzing the minute-frequency in three certain years.
These results enlarge the analogies between the turbulence and financial market dynamics, and provide a profound understanding beyond the phenomenological description of
multifractal scaling in stock price fluctuations, thus may help us to well
model the dynamic evolution of financial markets based on multiplication cascade
process.

\section*{Acknowledgments}
The authors acknowledge the support of the National Natural Science
Foundation of China (Grant Nos. 10975126, 91024026),
the Major Important Project Fund for Anhui University Nature
Science Research (Grant No. KJ2011ZD07), and the
specialized Research Fund for the Doctoral Program of Higher
Education of China(Grant No.£º20093402110032)

\end{document}